**Nanorippling of GaAs (001) surface near the threshold energy of sputtering at normal ion incidence**

Debasree Chowdhury, Debabrata Ghose[*], Safiul Alam Mollick, Biswarup Satpati and Satya Ranjan Bhattacharyya

*Saha Institute of Nuclear Physics, Sector – I, Block – AF, Bidhan Nagar, Kolkata 700064, India*

**Abstract**

Ripple formation driven by Ehrlich-Schwoebel barrier is evidenced for normal incidence 30 eV $Ar^+$ bombardment of GaAs (001) surface at elevated target temperature. The pattern follows the symmetry of the bombarded crystal surface. The results can be described by a non-linear continuum equation based on biased diffusion of adspecies created by ion impact.

Self-organized processes far from equilibrium have been considered as a promising route for the preparation of a rich variety of surface patterns at the nanometer length scale. One of the possibilities of this kind is the spontaneous pattern formation during thin film growth by molecular beam epitaxy (MBE) [1 - 4], where a number of investigations were dealt with technologically important GaAs (100) semiconductor material because of its unique optical and electrical properties [5]. For the case of homoepitaxial growth of GaAs, one observes the formation of elongated mound structure orientated along the $<1\bar{1}0>$ direction [1, 2]. This type of surface instability is related to the effect of Ehrlich-Schwoebel (ES) barrier [6, 7] which inhibits the downward movement of adatoms at surface step edges.

In this letter, we will show for the first time that similar type of ES instability can occur on GaAs (001) surface at elevated target temperature by ultralow energy $Ar^+$ ion irradiation at normal ion incidence. In this case, an array of nanowires resembling ripple-like structure develops on the surface. Such ordered ripples were not found in the case of deposition on GaAs crystal [1, 2]. We have chosen a beam of energy 30 eV, which is very close to the reported value of threshold energy ($\approx$ 20 eV) for sputtering of GaAs [8]. The incident ion beam breaks the bonding of GaAs to free As atoms which tend to vaporize from the surface because of its high vapour pressure, thus establishing a Ga enriched surface. The displaced lattice atoms remain mostly as surface adatoms as their energies are much less than the surface binding energy (~ 5 eV [9]) which the atoms must overcome before leaving the surface (sputtering). This means that at such low bombarding energy the sputter-yield is negligibly small, while the adatom yield is significantly higher. This phenomenon is verified experimentally [10] and also shown theoretically by molecular dynamics simulation [11]. The surface of a semiconductor material is known to become amorphous, in contrast to metallic targets, when bombarded by energetic ions at room temperature. Thus, a high sample temperature is needed to keep the surface crystalline during ion irradiation. The recrystallization temperature of GaAs is measured to be about 300 $^0$C [12]. When the ion bombardment is carried out above the recrystallization temperature, the bulk defects are dynamically annealed and the surface defects, i.e. adatoms, vacancies and their clusters tend to self-organize to form the nanostructures following the step edge dynamics of the crystalline surface.

The GaAs (001) wafers with a miscut of $\pm\ 0.5^0$ were bombarded with 30 eV Ar$^+$ ions at normal ion incidence using a broad ion beam extracted from an inductively coupled RF discharge ion source equipped with three graphite grid ion optical system (M/s Roth & Rau Microsystems GmbH, Germany) [13]. The beam current was 40 µA cm$^{-2}$. The sample surface was heated to $450^0$ C by a radiation heater from the front side. The surface morphology of the irradiated samples was examined *ex-situ* by atomic force microscopy (AFM) and the microstructural analysis of the surface pattern was done by cross-sectional transmission electron microscopy (XTEM).

A series of samples was irradiated with different fluences ranging from $1 \times 10^{17}$ to $7 \times 10^{19}$ ions cm$^{-2}$ and at the temperature of $450^0$ C. Figure 1 shows the corresponding AFM images exhibiting the evolution of the surface topography. At the early stages of irradiation, poorly ordered short ripples are formed. The ridges of the ripple are found to align along the $<1\bar{1}0>$ direction. With increasing fluence, the ripples become longer and more regular. The ripple pattern, for lower fluences, exhibits a large number of topological defects in the form of abrupt termination, lateral merging and Y-like junctions. The number of defects, however, is greatly reduced at high fluence irradiation with the improvement of the order of the structure. This can be recognized from the two-dimensional Fourier transform (FFT) of the AFM images. Initially a wide range of spatial frequencies is found to concentrate in an elliptic domain with its major axis parallel to the <110> direction. Subsequently, with the bombardment time the frequency domain becomes narrower with the development of a two-lobe symmetric pattern indicating the selection of a narrow band of wavelength. Finally, the appearance of higher order diffraction spots on either side of the central bright spot at larger fluences reveals the formation of a very regular ripple pattern.

The wavelength of the ripples can be determined from the one dimensional structure factor along the fast scan direction: $S(k,t) = \langle \hat{h}(k,t)\, \hat{h}(-k,t) \rangle$, where $\hat{h}(k,t)$ is the Fourier transform of the surface height $h$ and $k$ is the spatial frequency. The rms surface roughness is calculated from the relation $w = \sqrt{\langle [h(\mathbf{r},t) - \langle h(\mathbf{r},t) \rangle]^2 \rangle}$, where **r** is the lateral separation between two surface positions. Figures 2(a) and (b) show the variations of *rms* surface roughness $w$ and the wave length $l$ of the ripples as a function of ion fluence. Both the quantities exhibit a similar type of variation, i.e. after an initial increase at low fluence, $w$ as well as $l$ saturates at high fluence. From a power law fitting to the increasing part of the temporal evolution of the ripple amplitude and wavelength, the growth exponent ($\beta$) and the dynamic exponent ($1/z$) can be calculated. The values are, respectively, $\beta = 0.45 \pm 0.01$ and $1/z = 0.27 \pm 0.01$. We have also measured the ripple facet angle from the one dimensional angle distribution along the direction of the wave vector. The facet angle increases with fluence until it saturates to a value of $\sim 9^0$ at the highest fluence of $2.5 \times 10^{19}$ ions cm$^{-2}$ (Fig. 2(c)).

Figure 3 depicts the XTEM image of the irradiated surface at temperature $450^0$ C. A symmetric saw tooth like structure is visible without capping of amorphous layer. The pattern is highly crystalline with the same orientation as the original GaAs crystal. In another experiment with x-ray photoelectron spectroscopy, we find a change in surface stoichiometry with an excess of 15 at.% of Ga.

Since the morphological behavior resembles to the self-organized growth of GaAs during MBE [1, 2] a similar mechanism pertaining to homoepitaxial growth might be applicable here. The principal source of anisotropy in the developed structure is the ES biased intra- and interlayer mass transport. The usual theoretical approach to this type of growth phenomena is represented by a continuum theory, in which the local surface height $h(x, y, t)$ obeys the following conservation law [14]:

$$\frac{\partial h}{\partial t} + \nabla \cdot j = 0, \qquad (1)$$

where $j$ is the surface currents due to diffusion and can be divided into two components, namely, the classical Herring-Mullins (HM) diffusion current [15] $j_{HM} \sim \nabla(\nabla^2 h)$, and the nonequilibrium slope-dependent current $j_{ES}(m_{x,y} = \partial_{x,y} h)$, due to ES instabilities. For the latter, the following form of surface current is suggested $j_{ES} \sim m_{x,y}(1 - \delta m_{x,y}^2)$, where the first term refers to the destabilized uphill current and the second term corresponds to a downward current [16 - 18]. The growth front selects a stable slope $m_0 = \pm\sqrt{1/\delta}$ where the net ES current is zero. In the present experiment, we measure $m_0 \approx tan9^0$, so the value of the parameter $\delta$ comes out as 40. Finally, adding a nonlinear surface current of Kadar-Parisi-Zhang (KPZ) type, Eq. (1) yields

$$\frac{\partial h}{\partial t} = -\sum_{i=x,y} \varepsilon_i [\partial_i h(1 - \delta(\partial_i h)^2)] + \sum_{i,j=x,y} [-K_{i,j} \partial_i^2 \partial_j^2 h - \lambda_{i,j} \partial_i^2 (\partial_j h)^2], \qquad (2)$$

where ε, $K$ and $\lambda$ are positive coefficients and related to ES diffusion, HM diffusion and nonlinear KPZ terms, respectively. Here a conserved KPZ term is introduced in order to allow conservation of the total number of particles on the surface [19]. This term is known to induce ripple coarsening [20]. The above equation is similar to that derived by Ou et al. [18] for isotropic ES diffusion currents.

The GaAs (001) surface has a rectangular symmetry characterized by two mutually perpendicular directions <1$\bar{1}$0> and <110> [21]. The two uphill currents, $j_{<1\bar{1}0>}$ and $j_{<110>}$ are not equivalent in terms of ES barrier heights and activation energies of surface diffusion [21-24]. For Ga adatoms, a finite ES barrier exists along <110> but not along <1$\bar{1}$0> [21]. On the other hand, there is no ES barrier for As adatoms [21]. Thus intra-layer diffusion of Ga along the direction <1$\bar{1}$0> is possible, but the interlayer diffusion along <110> is blocked by the ES barrier, thereby resulting in an uphill current perpendicular to the surface. The measured growth exponent $\beta = 0.45$ suggests a barrier height around 0.1 eV [25] which is consistent to the value used in the simulation of reentrant mound formation in GaAs (001) homoepitaxy [1].

Equation (2) has been numerically integrated on a grid consisting of 256 × 256 lateral nodes with $\Delta x = \Delta y = 1$ and $\Delta t = 0.01$. Here $x$ and $y$ correspond to the crystallographic directions <110> and <1$\bar{1}$0>, respectively. The parameter values chosen are: $\varepsilon_x = 1$, $\varepsilon_y = 0.01$, $K_{i,j} = 1$, $\lambda_{i,x} = -1$ and $\lambda_{i,y} = -0.1$. The computed three-dimensional morphologies corresponding to different number of iteration (equivalent to different irradiation fluence) are shown in Fig. 4. The results are qualitatively similar to that obtained by AFM measurements (cf., Fig. 1), i.e. the amplitude and wavelength of the ripples increase in course of time, but there is no tendency of ultimate saturation for long time bombardment. It should, however, be noted that the present continuum equation produces a large band of unstable wave vectors contrary to the experimentally observed small region of unstable wave vectors. In order to explain this, Bradley and Shipman [26] suggested coupled continuum equations between the topography and the compositional change of the upper surface layer due to ion irradiation. Unfortunately, their theory is valid for an amorphous surface where the ES barrier is nonexistence and therefore, cannot be applied here as such.

In summary, we observe the development of nano-ripples on GaAs (001) surface near the threshold energy for sputtering at elevated temperature. The pattern formation is analogous to that observed in homoepitaxial growth by MBE, but it has greater order and regularity than that obtained by MBE. It is thought that such pattern is generated by the diffusion of ion-induced-adspecies influenced by ES barrier. A nonlinear anisotropic continuum equation accounting the ES instability can explain well the evolution of the ripple structure.

The authors are grateful to Prof. R. Mark Bradley of Colorado State University, USA for many useful discussions in course of the work.

**Figures:**

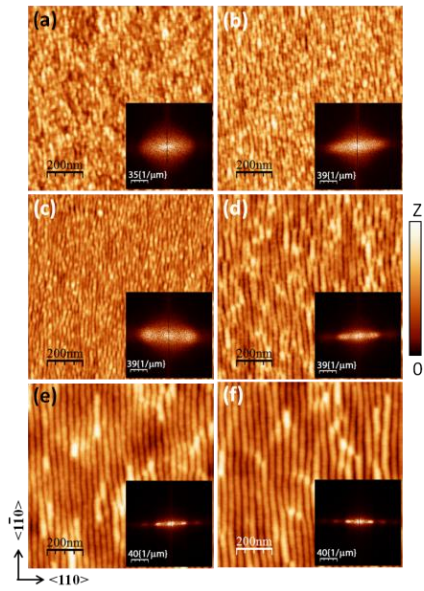

Fig. 1. The AFM topography and the corresponding FFT images of 30 eV Ar$^+$ irradiated GaAs (001) surfaces at different ion fluences: (a) $1 \times 10^{17}$ ions/cm$^2$; (b) $2.2 \times 10^{17}$ ions/cm$^2$; (c) $5 \times 10^{17}$ ions/cm$^2$; (d) $2.2 \times 10^{18}$ ions/cm$^2$; (e) $2.3 \times 10^{19}$ ions/cm$^2$; (f) $7 \times 10^{19}$ ions/cm$^2$.

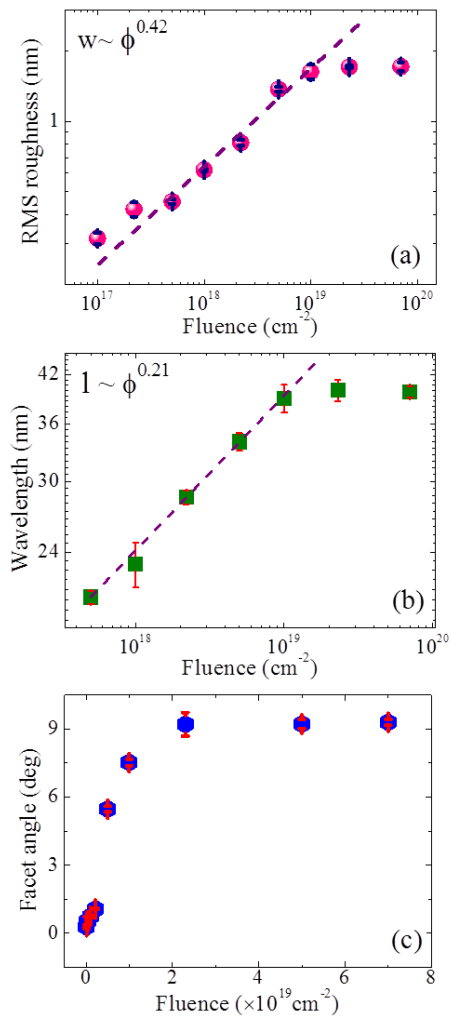

Fig. 2. (a) the rms roughness versus ion fluence, (b) the wavelength versus ion fluence and (c) the facet angle versus ion fluence.

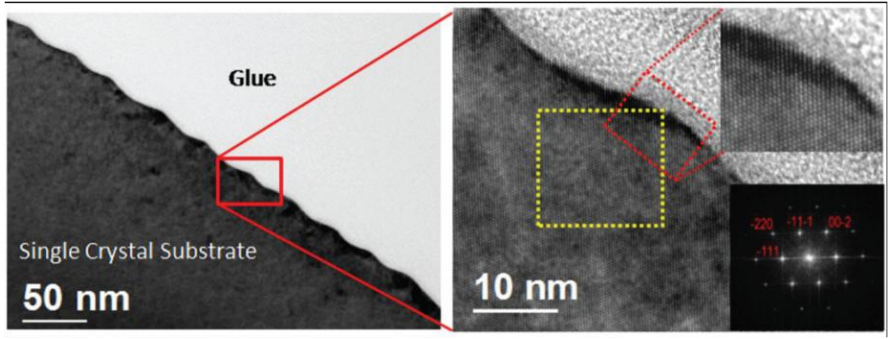

Fig. 3. Cross sectional TEM image of the rippled GaAs surface at the fluence $7 \times 10^{19}$ ions/cm$^2$ and temperature $450^0$ C. The dark part shows crystalline GaAs and the lighter part is the glue.

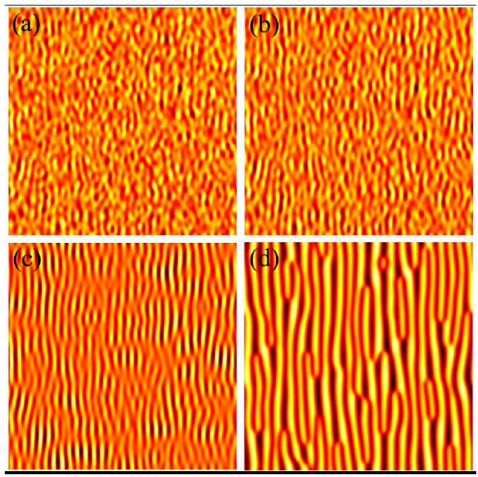

Fig. 4. Ripple pattern obtained by numerical integration of the continuum equation (2) at different integration steps (a) , (b), (c).